\documentstyle[12pt]{article}

\begin{document}

\begin{titlepage}

\begin{flushright}
\end{flushright}

\baselineskip 24pt

\begin{center}

{\Large {\bf Flavour-Changing Neutral Currents in the Dualized
Standard Model }}\\

\vspace{.5cm}

\baselineskip 14pt
{\large Jos\'e BORDES}\\
bordes\,@\,evalvx.ific.uv.es\\
{\it Departament Fisica Teorica, Universitat de Valencia,\\
  calle Dr. Moliner 50, E-46100 Burjassot (Valencia), Spain}\\
\vspace{.2cm}
{\large CHAN Hong-Mo, Jacqueline FARIDANI}\\
chanhm\,@\,v2.rl.ac.uk  \,\,\,  faridani\,@\,hephp1.rl.ac.uk \\
{\it Rutherford Appleton Laboratory,\\
  Chilton, Didcot, Oxon, OX11 0QX, United Kingdom}\\
\vspace{.2cm}
{\large Jakov PFAUDLER}\\
jakov\,@\,thphys.ox.ac.uk\\
{\it Dept. of Physics, Theoretical Physics, University of Oxford,\\
  1 Keble Road, Oxford, OX1 3NP, United Kingdom}\\
\vspace{.2cm}
{\large TSOU Sheung Tsun}\\
tsou\,@\,maths.ox.ac.uk\\
{\it Mathematical Institute, University of Oxford,\\
  24-29 St. Giles', Oxford, OX1 3LB, United Kingdom}
\end{center}

\vspace{.3cm}

\begin{abstract}
The Dualized Standard Model which gives explanations for both fermion
generations and Higgs fields has already been used to calculate fermion 
mass and mixing parameters with success.  In this paper, we extend its
application to low energy FCNC effects deriving bounds for various 
processes in terms of one single mass scale.  Using then experimental 
information from $K_L - K_S$ mass difference and air showers beyond the 
GZK cut-off, these bounds are converted into rough, order-of-magnitude 
predictions.  In particular, the estimates for the decay $K_L \rightarrow 
e^\pm \mu^\mp$ and for the mass difference between the neutral $D$-mesons 
seem accessible to experiment in the near future.
\end{abstract}

\end{titlepage}

\clearpage

\section{Introduction}

Based on a nonabelian generalization of electric-magnetic duality 
\cite{chantsou1} and a result on confinement of 't~Hooft \cite{thooft}, 
a Dualized Standard Model (DSM) \cite{chantsou2} was constructed which,
though remaining strictly within the standard model framework in 
contrast to most other attempts with a similar purpose \cite{ross},
yet offers an explanation for the existence of exactly 3 generations as 
broken dual colour and of Higgs fields as frame vectors in internal 
symmetry space.  The symmetry breaking pattern inherent in the scheme 
is such that at tree-level only the highest generation fermions have 
a mass (mass hierarchy) while the Cabibbo-Kobayashi-Maskawa (CKM)
matrix, whether for quarks or leptons, is the identity matrix.  Loop 
corrections, however, give nonzero masses to the 2 lower generations 
and nonzero mixing both between the $U$- and $D$-type quarks and
between the charged leptons and neutrinos.  A calculation of these 
corrections to 1-loop level has already been performed 
\cite{ckm,nuos,features} giving very encouraging
results.  By adjusting effectively only 3 real parameters, we have
calculated the following 14 quantities: the 3 mixing parameters (except
the CP-violating phase) in the quark CKM matrix $V_{rs}$, the 3 analogous
parameters in the lepton CKM matrix $U_{rs}$, the 6 known fermion masses
$m_c, m_s, m_\mu, m_u, m_d, m_e$, as well as estimates for the masses 
of the lightest neutrino $m_{\nu_1}$ and the `right-handed' neutrino $B$.  
Of these 14 calculated quantities, 10 agree as well as can be expected 
with experiment (i.e.\ all except $m_u$ and probably also $U_{e2}$, 
which are not good, and $m_{\nu_1}$ and $B$, which are experimentally 
unknown.)

Given these encouraging results it is natural to seek further tests for
the DSM scheme in other directions.  One obvious area which may yield
interesting tests is in flavour-changing neutral current (FCNC)
effects.  The fact that generation has been identified to dual colour 
which is itself a (broken) gauge symmetry means that any fermion
carrying a dual colour or generation index will interact by exchanging
(electrically neutral) dual colour gauge bosons (dual gluons) leading 
thus to flavour-changing neutral currents, which will manifest
themselves in rare FCNC decays and in the mass differences between
for example, the neutral strange, charm and bottom mesons.  The purpose 
of the present article is the evaluation of these FCNC effects as 
predicted by the DSM scheme and their comparison with the existing 
bounds obtained from experiment. 

The predictions for the various FCNC effects in schemes of so-called 
`horizontal symmetries' \cite{neutralcurrent,buchmuller}, of which the
DSM is one, depend on the masses of the exchanged gauge bosons and their
couplings to the relevant fermion states.  What is special for the DSM, 
however, is that, in view both of the intrinsic structure built into the 
scheme and of the calculations already performed which are referred to
above, most of these quantities are now known.  First,
by virtue of the Dirac quantization condition \cite{chantsou3}:
\begin{equation}
g_{3(2)} \tilde{g}_{3(2)} = 4\pi, \;\;\; g_1 \tilde{g}_1 = 2\pi,
\label{couplings}
\end{equation}
the coupling strengths $\tilde{g}_i$ of the dual gauge bosons are derivable 
from the coupling strengths $g_i$ of the ordinary colour and electroweak 
gauge bosons measured in present experiments.  Secondly, the branching 
of these couplings $\tilde{g}_i$ into the various physical fermion 
states are given by the orientations of these physical states in generation 
or dual colour space, and these orientations are already determined in 
the calculation of CKM matrices \cite{ckm,nuos} mentioned above.  Finally, 
in tree-level approximation, the masses of the dual gauge bosons are 
given in terms of the vacuum expectation values of the dual colour Higgs 
fields, the ratios between which are among the parameters determined in 
the calculation \cite{ckm} by fitting the CKM matrix.  Thus the only 
remaining unknown among the quantities required is the mass scale of 
the dual gauge bosons which, for reasons explained in \cite{ckm}, is 
not restrained significantly by the calculation reported there.  That 
being the case, one can now calculate in the DSM scheme all FCNC effects 
in tree-level approximation in terms of this single scale parameter.  
Or, in other words, given the experimental bound on any one FCNC effect, 
one can evaluate the correlated bound on any other such effect.  

Now, it turns out that in the tree-level spectrum found for the dual 
gauge bosons, one particular state has a much lower mass than the rest, 
so that the calculation of low energy FCNC effects becomes quite simple, 
being dominated by just the exchange of this one boson.  In that case 
it happens that it is the mass difference between the $K_L$ and $K_S$ 
mesons which sets the tightest bound on the gauge boson mass. From 
this, one then obtains the correlated bounds on all other low energy 
FCNC effects.

At present, directly from the DSM scheme, one obtains only these bounds.
However, going outside the DSM scheme proper there is one valuable 
piece of information borrowed from cosmic ray physics which, if taken
seriously, converts the bounds mentioned above into order-of-magnitude
predictions.  This concerns the small number of air showers with primary 
energy in excess of $10^{20}$ eV which have been observed over the last 
decades \cite{boratav,auger}.  These events are a mystery in that protons, 
which are thought to account for high energy showers, react readily via 
radiative pion-production with the 2.7 K microwave background and quickly 
degrade in energy.  Indeed, according to Greisen, Zatsepin and Kuz'min 
\cite{greitsemin} the spectrum of protons should cut off sharply at about 
$5 \times 10^{19}$ eV, unless they originate at distances of less than 50 Mpc,
which is thought unlikely, there being no known nearby sources capable of
producing such high energy particles.

One suggestion \cite{airshower1,airshower2} was that these rare showers with 
energy above the GZK cut-off are due to neutrinos rather than protons. 
Neutrinos, being electrically neutral, could survive a long journey
through the 
2.7 K microwave background with their energy intact.  However, neutrinos with 
only their presently known weak interactions will not have a large enough
cross section with the air nucleus to produce air showers with the observed 
characteristics.  The neutrino explanation, therefore, is not feasible
unless, for some reason, neutrinos acquire a strong interaction at high 
energy.  Now, the beauty in the DSM scheme is that this is exactly what has
been predicted \cite{chantsou2}.  Neutrinos, like other fermions, occur 
in 3 generations, which are now identified with dual colour. This means that 
neutrinos will interact via the same exchanges of dual colour gauge bosons 
which give rise to the flavour-changing neutral currents considered above.  
Dual colour gauge bosons being massive, this new interaction for neutrinos 
will be suppressed at low energy like all FCNC effects and be at present
unobservable.  But at energies comparable to the dual colour gauge boson 
mass, it will come into its own, and is expected to be strong given that 
the dual colour coupling is constrained by the Dirac quantization condition 
\cite{chantsou3}.  Indeed, the analysis in \cite{airshower1} of this 
DSM neutrino hypothesis offers not only qualitative explanations for 
several outstanding mysteries surrounding post-GZK air showers but even 
some quantitative predictions testable by future experiments.\footnote{There
has appeared in the literature \cite{halzen} a claim that the DSM neutrino
explanation along with most other particle physics explanations for post-GZK 
showers do not work.  We find the arguments presented there, based only on 
first order perturbation theory, inadequate for such a sweeping claim.  
See \cite{airshower2} and section 4.2 of this paper for further discussion 
on this point.}

As far as low energy FCNC effects are concerned, the relevance of the above 
discussion on post-GZK air showers lies in its giving a rough upper bound 
on the dual colour gauge boson mass.  The centre of mass energy for a 
neutrino with primary $E \sim 10^{20}$ eV impinging on a proton at rest is 
around 500 TeV, which ought to be comparable to the mass of the gauge 
boson exchanged before the resulting interaction becomes appreciable.  
Interestingly, as we shall see, this estimate is not far from the lower 
bounds obtained from present experiment on low energy FCNC effects.  As
a result, the correlated bounds deduced directly from the DSM scheme 
above become now absolute predictions, although still rather unsure ones 
to be trusted only at a rough order of magnitude level, given that the
estimate for the scale from air showers is by necessity crude and affects
the predictions sensitively.  Even then, we think the predictions useful
for planning future experiments since, as we shall see, there are several 
effects predicted with values only a couple of orders of magnitude below 
the existing experimental bounds.

It should be stressed, however, that the treatment in this paper suffers 
from one serious weakness in that it relies on the tree-level spectrum 
of the dual gauge bosons.  The couplings involved being large, it is 
unclear how the spectrum will change under higher order corrections which, 
at the present stage of development of the DSM scheme, one does not know
how to estimate.  And if this change in spectrum turns out to be drastic,
then it can affect our predictions even qualitatively, as we shall elucidate.
We can only hope at present that this does not happen.

In section 2, we review briefly those features of the Dualized Standard
Model relevant to our present study, and in section 3 construct its low 
energy effective Lagrangian.  Its experimental consequences on low energy
FCNC effects are analysed in section 4.

\section{The Dualized Standard Model}

According to \cite{chantsou2}, to which the reader is referred for details,
the terms of the Lagrangian of interest to us here, including the Higgs 
potential and the couplings of the dual gauge bosons to the Higgs and 
fermions fields, can be cast into the form:
\begin{eqnarray}
& & L_{\phi,C}+L_{C,\psi}+V_{\phi}  
\nonumber \\
& = & 
\bar{\phi}
\left[
- \tilde{g}_3 {C}^{b}_{gauge,\mu} 
\frac{\lambda_b}{2} - \tilde{g}_1 \left(  n + \frac{1}{3} \right)
{B}_{gauge,\mu} \right]^2 \phi 
\nonumber \\
&  + & 
\bar{\psi}_L \left[- i \tilde{g}_3 {C}^{b}_{gauge,\mu} 
\frac{\lambda_b}{2} - i \tilde{g}_1 \left(  {n_L} + \frac{1}{3} \right)
{B}_{gauge,\mu}  \right] \psi_L 
\nonumber \\
&  - &
\mu \sum_{(a)}|\phi^{(a)}|^2 + 
\lambda \left[ \sum_{(a)}|\phi^{(a)}|^2 \right]^2 +
\kappa \sum_{(a) \neq (b)}|\bar{\phi}^{(a)} . \phi^{(b)}|^2,
\label{lag0}
\end{eqnarray}
with $\mu$, $\kappa$ and $\lambda$ all positive, $n$ and $n_L$ odd integers,
and $\tilde{g}_3$ and
$\tilde{g}_1$ the dual gauge couplings related to the usual colour 
coupling $g_3$ and weak hypercharge coupling $g_1$ by the Dirac 
quantization condition (\ref{couplings}).  The Higgs ($\phi^{(a)}$, 
$a=1,2,3$) and the left-handed fermion ($\psi_L$) fields are in the 
fundamental while the gauge boson fields ($C^{b}_{gauge,\mu}$, $b=1,\ldots,8$)
are in the adjoint representation of the dual colour gauge group
$\widetilde{SU}(3)$.  The field $B_{gauge,\mu}$, which we shall also 
denote by $C^0_{gauge,\mu}$, represents the $\widetilde{U}(1)$ gauge field 
of dual (weak) hypercharge.  The number $n_{L} + \frac{1}{3}$ is the dual 
hypercharge of the multiplet $\psi_{L}$.  Note that we take a vanishing 
dual hypercharge for the right-handed fermion components in order to 
allow for a Majorana mass term for neutrinos in the Lagrangian \cite{nuos}. 
With this choice of hypercharges only the left-handed component contributes 
to the current in (\ref{lag0}) giving a $\;V-A\;$ fermion-dual boson 
interaction term.  

By virtue of the Higgs potential in (\ref{lag0}) the Higgs fields acquire 
nonzero vacuum expectation values which, because of the symmetries of the
Higgs potential, we are allowed to assign the following values:
\begin{eqnarray}
\phi^{(1)} = \zeta \left(
\begin{array}{c}
 x \\ 0 \\ 0 
\end{array} \right), \,\,\,
\phi^{(2)} = \zeta \left(
\begin{array}{c}
 0 \\ y \\ 0 
\end{array} \right) , \,\,\,
\phi^{(3)} =  \zeta \left(
\begin{array}{c} 
0 \\ 0 \\ z 
\end{array} \right),
\label{vevs}
\end{eqnarray}
where we have normalized the three fields according to:
$$
x^2 + y^2 + z^2 = 1, \,\,\,\,\,\,
\zeta^2=\frac{\mu}{2 \lambda}.
$$
As a result, the symmetry $\widetilde{SU}(3) \times \widetilde{U}(1)$ is
completely broken, giving masses to all the dual bosons and to the fermions. 
The mass terms for the dual bosons added to the gauge boson-fermion 
interaction term, give the piece of the Lagrangian (\ref{lag0}) which 
is relevant for our purpose: 
\begin{eqnarray}
& & L_{mass}+L_{C,\psi} =  
\nonumber \\
& &  C^{b}_{gauge,\mu} M_{b,b'} C^{b'}_{gauge,\mu} + 
\bar{\psi}_L \left[- i \tilde{g}_3 C^{b}_{gauge,\mu} 
\frac{\lambda_b}{2} - i \tilde{g}_1 \left( n_L + \frac{1}{3} \right)
C^0_{gauge,\mu}  \right] \psi_L. 
\nonumber \\
\label{lag1}
\end{eqnarray}

The $9 \times 9$ mass matrix $M_{b,b'}$ has a piece which is already 
diagonal in the original gauge basis, namely that corresponding to the 
bosons $C^b_{gauge}$ with $b=1,2,4,5,6,7$, which are thus already the 
physical mass eigenstates at tree-level.  These have the masses:
\begin{eqnarray}
M_b & = & \zeta \frac{\tilde{g_3}}{2} \sqrt{x^2 + y^2}, \, \, \, \, \, \, 
 b=1,2;
 \nonumber \\
M_b & = & \zeta \frac{\tilde{g_3}}{2} \sqrt{z^2 + x^2}, \, \, \,\, \, \, 
 b=4,5;
 \nonumber \\
M_b & = & \zeta \frac{\tilde{g_3}}{2} \sqrt{y^2 + z^2}, \, \, \,  \, \, \, 
 b=6,7; 
\label{mass0}
\end{eqnarray}
The remaining dual bosons ($C^b$ with $b=3,8,0$), since they are diagonal 
in dual colour, mix with one another in the mass matrix:
\begin{eqnarray}
M^2 = \zeta^2
\left(
\begin{array}{ccc}
 \frac{\tilde{g_3}^2}{4} (x^2 + y^2) &
 \frac{\tilde{g_3}^2}{4 \sqrt{3}} (x^2 - y^2) &  
 - \frac{\tilde{g_3} \tilde{g_1}}{3} (x^2 - y^2) \\ 
 \frac{\tilde{g_3}^2}{4 \sqrt{3}} (x^2 - y^2)  & 
 \frac{\tilde{g_3}^2}{12} (x^2 + y^2 + 4 z^2)  &
 -\frac{\tilde{g_3}\tilde{g_1}}{3 \sqrt{3}} (x^2 + y^2 - 2 z^2)  \\
 -\frac{\tilde{g_3}\tilde{g_1}}{3} (x^2 - y^2)  & 
 -\frac{\tilde{g_3}\tilde{g_1}}{3 \sqrt{3}} (x^2 + y^2 - 2 z^2)  & 
 \frac{4 \tilde{g_1}^2}{9} (x^2 + y^2 + z^2)  
\end{array}
\right),
\nonumber \\
\label{mass1}
\end{eqnarray}
where the ordering $3,8,0$ in the matrix elements has been followed.

We see thus that at tree-level the masses of the dual gauge bosons are
given in terms of the vacuum expectation values of the Higgs fields, namely
$x,y,z$.  These parameters are, in principle, unknown but were determined
in our previous paper \cite{ckm} by fitting the experimental quark CKM 
matrix and the masses of the two heavier generations of quarks and charged 
leptons, giving: 
\begin{equation}
x=1, \,\,\, y=5 \times 10^{-5}, \,\,\, z=1 \times 10^{-8}.
\label{vevs1}
\end{equation}
The manner in which this was done, together with the whole interesting question
of the fermion mass matrices, is described in detail in \cite{ckm} and
will not be repeated here.  Notice, however, that the parameter $\zeta$ 
which gives the mass scale is still unconstrained.

In addition, the above fit gave the rotation matrices transforming from 
the original gauge basis to the basis of the physical states for each
fermion-type.  We give here the transformation matrices for quarks and 
charged leptons which will be relevant later for the comparison to 
experimental data.  With the notation:
\begin{equation}
\psi_{gauge,L}^A = S^A \psi_{physical,L}^A
\label{rotationf}
\end{equation}
these rotation matrices are\footnote{We arrange the matrix elements here
in order of decreasing fermion mass in accordance with the convention
adopted in \cite{ckm}.}:
\begin{eqnarray}
& & 
S^U=
\left(
\begin{array}{ccc}
0.9999 & -0.0127 & 0.0045 \\
0.0135 & 0.9163 & -0.4002 \\
0.0009 & 0.4002 & 0.9164 
\end{array}
\right),
\nonumber \\
& & 
S^D=
\left(
\begin{array}{ccc}
0.9983 & -0.0545 & 0.0215 \\
0.0566 & 0.8044 & -0.5914 \\
0.0149 & 0.5916 & 0.8061 
\end{array}
\right), 
\nonumber \\
 & & 
 S^L=
\left(
\begin{array}{ccc}
0.9961 & -0.0829 & 0.0293 \\
0.0829 & 0.7740 & -0.6277 \\
0.0294 & 0.6277 & 0.7779 
\end{array}
\right).
\label{mixingf}
\end{eqnarray}

Notice that the CKM matrix is given in terms of these
matrices as the combination $(S^U)^\dagger S^D$ and it was found to be in 
excellent agreement with the present experimental values \cite{ckm}.

\section{One Dual Gluon Exchange and FCNC}

The dual colour gauge bosons (or dual gluons) are responsible for transitions 
between fermions of different generations but of the same type.  Since the
dual gluons carry no electric charge, this is equivalent to the existence 
of neutral currents with changes of flavour (FCNC).  In this section we 
develop the formalism needed to study their phenomenological implications,
the actual analysis of which, however, is postponed to the next section. 
We study in detail in this paper only one dual gluon exchange.  As a
check, we shall make estimates of the contributions from higher order
corrections.  However, the couplings being large by virtue of the Dirac
quantization condition (\ref{couplings}), we cannot really be confident 
in perturbative calculations beyond the tree-level.  For that reason, 
the results we shall present in the next sections have to be regarded 
only as rough, order-of-magnitude, estimates. 

The relevant term of the interaction Lagrangian is given in the usual 
formalism by:
\begin{equation}
L= -i \sum_A \left[
\tilde{g_3} J_{L,\mu}^{A,a} C^{\mu}_{gauge,a} +
\tilde{g_1} J_{L,\mu}^{A,0} C^{0,\mu}_{gauge} \right]. 
\label{lag3}
\end{equation}
The current is expressed using a triplet of left-handed 
gauge fermion fields and the sum over the index $A$ runs over the different 
fermion-types ($U, D, L, N$).  Explicitly it is given in terms of these fields 
and the Gell-Mann matrices by:
\begin{eqnarray}
J_{L,\mu}^{A,a} & = & \bar{\psi}_L^A \gamma_\mu \frac{\lambda^a}{2} \psi^A_L, 
\nonumber \\
J_{L,\mu}^{A,0} & = & \bar{\psi}_{L}^A \gamma_\mu 
\left(
n_{L}^A +
\frac{1}{3}
\right)
 I \psi^A_{L}.
\label{current0}
\end{eqnarray}
Both the fermion and dual gauge boson fields are here given in the 
(gauge) basis where the vacuum expectation values of the Higgs fields 
appear as in (\ref{vevs}).  We need now to express them in terms of 
their physical states.  

The mixing in the fermion sectors was discussed in section 2, and when
the fermion fields are written in the physical basis, the 
current takes the form\footnote{In the following we will suppress  the 
index $physical$ in the fermionic fields.}:
\begin{equation}
J_{L,\mu}^{A,a}  =  \bar{\psi}_L^A \gamma_\mu 
(S^A)^\dagger \frac{\lambda^a}{2} S^A \psi^A_L.
\label{current1} \\
\end{equation}
It is worth noting that the unitary matrices $S^A$, being a transformation 
in generation space, do not commute with the Gell-Mann matrices. As a 
consequence, one cannot rotate away, even with massless neutrinos, the 
transformation in the leptonic sector and this has observable consequences. 
This fact is common to many models with horizontal symmetries.

Next, consider the physical states in the gauge boson sector.  At tree-level
these are given in (\ref{mass0}) and by the diagonalization of the matrix
(\ref{mass1}).  In the situation when the v.e.v.'s of the Higgs fields in
(\ref{vevs}) are hierarchical, namely $x \gg y \gg z$, which is indeed the
case in the fit (\ref{vevs1}) obtained in \cite{ckm}, the matrix (\ref{mass1})
can be diagonalized algebraically giving the eigenvalues as:
\begin{eqnarray}
& & 
M^{physical}_1= \frac{2 \tilde{g}_1}{\left[1+
\frac{16 \tilde{g}_1^2}{3 \tilde{g}_3^2} \right]^{1/2}} 
\, \, \zeta z, 
\nonumber \\
& &
M^{physical}_2 = \frac{\tilde{g}_3}{2}
\left[ 
\frac{1+\frac{16 \tilde{g}_1^2}{3 \tilde{g}_3^2}}
{1 + \frac{4 \tilde{g}_1^2}{3 \tilde{g}_3^2}} \right]^{1/2} \zeta y, 
\nonumber \\
& & 
M^{physical}_3 =  \frac{ \tilde{g}_3}{\sqrt{3}}
\left[ 1+\frac{4 \tilde{g}_1^2}{3 \tilde{g}_3^2} \right]^{1/2} \zeta x.
\label{mass2}
\end{eqnarray}
Comparing this result with the masses previously found (\ref{mass0}), we 
see that there is one boson ($M^{physical}_1$) much lighter than the others,
the next lightest having, according to (\ref{vevs}), (\ref{mass0}) and 
(\ref{mass2}), masses some 2 orders of magnitude heavier.  Hence, given 
that the FCNC effects we seek depend on the mass of the gauge boson 
exchanged as $M^{-4}$, one sees that in the sum over the boson fields 
in equation (\ref{lag3}), one needs to keep only the contribution of 
this lightest boson.  The corresponding physical field is given just 
by the dual boson $C^8_{gauge,\mu}$ with a small admixture of the dual 
hypercharge boson, namely:
\begin{equation}
|C_{physical}^1 \rangle =
\frac{1}{k_1}
\left[ |C_{gauge}^8 \rangle + \frac{\sqrt{3}\tilde{g_3}}{4 \tilde{g_1}}
|C_{gauge}^0 \rangle \right],
\label{mixingb0}
\end{equation}
where $k_1$ is a normalization factor.

For completeness, we give the full diagonalizing matrix for the matrix 
(\ref{mass1}) in the limit $x \gg y \gg z$, with the convention 
$M^2_D=U^\dagger M^2 U$:
\begin{eqnarray}
U =
\left(
\begin{array}{ccc}
0   & -\frac{1}{k_2} \frac{1}{\sqrt{3}} \left(1+\frac{16 \tilde{g}_1^2}{3\tilde{g}_3^2}
\right) & \frac{1}{k_3} \\
\frac{1}{k_1} & \frac{1}{k_2} & \frac{1}{\sqrt{3} k_3} \\
\frac{1}{k_1} \frac{\sqrt{3} \tilde{g}_3}{4 \tilde{g}_1}&
 - \frac{1}{k_2} \frac{4 \tilde{g}_1}{\sqrt{3} \tilde{g}_3} &
 - \frac{1}{k_3} \frac{4 \tilde{g}_1}{3 \tilde{g}_3}
\end{array}
\right),
\label{mixingb1} 
\end{eqnarray}
where $k_i$ are the normalization constants of the respective column vectors.

After performing the rotation to the physical bases in both the fermionic 
and bosonic sectors, we end up with the interaction Lagrangian which we 
write as:
\begin{equation}
L= -i \sum_{A} 
\bar{\psi}^A_L \gamma_\mu g^{A,b} 
\psi^A_L C^\mu_{physical,b} 
\label{lag4}
\end{equation}
where
\begin{equation}
g^{A}_b  =  (S^A)^\dagger \left[
\tilde{g_3} \frac{\lambda_{8}}{2} U_{8,b} +
\tilde{g}_1 (n^A_L +
\frac{1}{3}) U_{0,b} I \right] S^A  ,
\label{coupling1}
\end{equation}
is a matrix in generation space, with the index $A$ taking the values 
$U,D,L,N$ and the index $b$ running over $1,2,3$.  As we shall
see, as far as processes with change of flavour is concerned, the second 
term in (\ref{coupling1}) dependent on the dual hypercharge, and hence 
proportional to the identity matrix in generation space, does not contribute.  

Next we change to the language of effective Lagrangians to study further
the flavour changing effects induced by (\ref{lag4}).  The effective 
Lagrangian will have the current$\times$current structure and will give, 
in first order, the same result as the interaction Lagrangian (\ref{lag3}) 
in second order when the masses of the intermediate bosons are much heavier 
than any other energy scale involved in the process.  Although the two
formalisms are equivalent, the effective Lagrangian language is more c
onvenient 
when considering low energy effects of a theory broken at a high scale 
\cite{buchmuller}.
 
Since only one intermediate boson (\ref{mixingb0}) need to be considered, 
the effective Lagrangian is proportional to the inverse square of the 
lowest Higgs v.e.v. $\zeta z$:
\begin{equation}
L_{eff}=
\frac{1}{2 \zeta^2 z^2} \sum_{A,B} (J^{\mu}_A)^{\dagger} J_{\mu,B}.
\label{lageff}
\end{equation}
Equivalently, if preferred, it can be written in terms of an effective 
coupling and the mass of the exchanged gauge boson:
\begin{equation}
\frac{g_{eff}^2}{M^2} = \frac{1}{\zeta^2 z^2},
\label{strength}
\end{equation}
where
\begin{equation}
g_{eff}^2 = \frac{3}{4} \frac{\tilde{g}_3^2}
{1 + \frac{3}{16}\frac{\tilde{g}_3^2}{\tilde{g}_1^2}}.
\label{geff}
\end{equation}
The fermionic current in (\ref{lageff}) has the same form as in (\ref{lag4}) 
but contains now only the hypercharge component proportional to the identity 
matrix and the component proportional to the Gell-Mann matrix $\lambda_8$.
Explicitly, the couplings are given as: 
\begin{equation}
g^{A}  =  (S^A)^\dagger \left[
\frac{1}{\sqrt{3}} \lambda_{8} + \frac{1}{2} ( n^A_L + 
\frac{1}{3}) I \right] S^A, 
\label{coupling3}
\end{equation}
from which the explicit dependence of the original gauge couplings
$\tilde{g}_i$ cancel.   

We notice in (\ref{coupling3}) that although the quantity inside the
brackets is diagonal in the gauge basis, it is not diagonal in
the physical fermion basis since $\lambda_8$ does not commute with the
fermion mixing matrices $S^A$.  This means that (\ref{lageff}) will
give transitions between physical fermion states of different flavours
but of the same type.  In order words it will give rise to the so-called 
flavour-changing neutral current effects that we seek.  It will give
transitions, of course, also between physical states with the same
flavours, but we shall not consider here such `diagonal processes' since
they appear also in the Standard Model and, at present, are not likely 
to give experimentally distinctive signatures.  We shall therefore
extract from (\ref{lageff}) only that piece relevant for flavour-changing
transitions.  The result will then be just the product of, firstly, 2 
fermionic currents each one made out of single left-handed fermion fields 
with a definite physical flavour, secondly, the effective coupling strength 
which is essentially the inverse square of the Higgs v.e.v $\zeta z$, 
and, thirdly, a group factor ($f^{A,B}_{\alpha,\beta;\alpha',\beta'}$)
coming from the rotation between gauge and physical states in generation
or dual colour space.  Explicitly, the final form of the effective 
Lagrangian which we shall use for the phenomenological analysis in the
next section is:
\begin{equation}
L_{eff}=
\frac{1}{2 \zeta^2 z^2} 
\sum_{A,B} f^{A,B}_{\alpha,\beta;\alpha',\beta'} 
(J^{\mu \dagger}_A)^{\alpha,\beta}  (J_{\mu,B})^{\alpha',\beta'}\,,
\label{laeff1}
\end{equation}
with currents of the usual $\;V-A\;$ form:
\begin{equation}
(J^{A}_\mu)_{\alpha,\beta} = \bar{\psi}_{L,\alpha}^A \gamma_\mu \psi_{L,\beta}^A,
\label{current2}
\end{equation}
and the group factor reduced to:
\begin{equation}
f^{A,B}_{\alpha,\beta;\alpha',\beta'} =
S^{A*}_{3,\alpha} S^{A}_{3,\beta}
S^{B*}_{3,\alpha'} S^{B}_{3,\beta'} 
\label{groupfactor1}
\end{equation}
The simple structure of (\ref{laeff1}) comes from the fact that the exchange 
of only one dual gauge boson need be considered.  In the formula for the
group factor (\ref{groupfactor1}), we note that for some of the leptonic
decays of mesons to be considered in the next section, there are in general
additional terms diagonal in one of the fermion vertices which depend on
the dual weak hypercharges of the left-handed fermions.  However, for 
the minimal choice $-2/3$ advocated in \cite{chantsou2} corresponding to 
$n_L^A = -1$ in (\ref{coupling3}), these terms cancel.

Notice that the group factors $f^{A,B}_{\alpha,\beta;\alpha',\beta'}$
are given entirely in terms of the matrices $S^A_{\alpha,\beta}$ of
(\ref{mixingf}) determined in our earlier work \cite{ckm} from fitting the 
CKM matrix, so that the only remaining free parameter in (\ref{laeff1}) is 
the mass scale $\zeta z$ which has yet to be estimated from experiment.

\section{Phenomenological Implications}

In this section we apply the formalism developed above to phenomenology.
Our first concern is to deduce bounds on the remaining unknown parameter
$\zeta z$.  As explained in the introduction, a lower bound will come
from low energy flavour-changing neutral current effects, while an upper
bound will arise in high energy air showers from cosmic rays.  It turns
out that the tightest lower bound one can obtain at present comes from
the $K_L-K_S$ mass difference, which is what we shall now consider.

\subsection{$K_L - K_S$ Mass Difference}

The relevant piece of the effective Lagrangian (\ref{lageff}) contributing
to the mass difference in the $K_L-K_S$ system is:
\begin{equation}
L^{D,D}_{eff} = \frac{1}{2 \zeta^2 z^2} 
|f^{D,D}_{2,3;2,3}|
\left( \bar{s}_L \gamma^\mu d_L \right)
\left( \bar{s}_L \gamma_\mu d_{L} \right),
\label{masskaon}
\end{equation}
which gives a contribution to the mass-splitting from dual gluon exchange 
of the form:
\begin{equation}
\Delta m_K =
\frac{1}{\zeta^2 \,\, z^2} |f^{D,D}_{2,3;2,3}|
\langle K^0| \left[\bar{s}_L \gamma^\mu d_L \right]^2 |\bar{K}^0 \rangle.
\end{equation}
Evaluating the matrix elements in the vacuum insertion approximation one 
arrives at the following expression for the mass-splitting:
\begin{equation}
\Delta m_K =
\frac{1}{\zeta^2 \,\, z^2} |f^{D,D}_{2,3;2,3}|\frac{f_K^2 \, \, m_K}{3}
\label{delmK}
\end{equation}

Taking $f_K^2=1.4 \times 10^{-8}\mbox{ TeV}^2$ for the $K$ decay constant, 
and $m_K=498$ MeV for the $K$ mass, a direct comparison of (\ref{delmK})
to the experimental value\footnote{All experimental values of rare decay
branching ratios and mass differences used in the analysis of this section 
are taken from \cite{databook}.} $\Delta m_K(K_L-K_S) =3.5 \,\times \, 
10^{-12}\mbox{ MeV}$ gives a lower bound:
\begin{equation}
\zeta \,\, z \geq 400\mbox{ TeV}.
\label{bound}
\end{equation}

We could instead compare (\ref{delmK}) with the Standard Model contribution 
of the charm quark to the $K_L - K_S$ mass splitting as advocated by
\cite{neutralcurrent}.  Using then the Gaillard-Lee \cite{gaillardlee}
effective Lagrangian for the charm contribution, one obtains that the 
bound on $\zeta z$ is increased up to 600 TeV.  On the other hand, it is 
by now generally accepted that the vacuum insertion approximation used in
deriving the above bounds overestimates the matrix element and ought to
be reduced by a factor of the order of 0.7 \cite{what}, in which case the
bound on $\zeta z$ can be lowered down to around 300 TeV.  To be specific,
the estimates in this paper are calculated using the bound (\ref{bound}),
but the corresponding estimates for other values of $\zeta z$ are easily
deduced given their $(\zeta z)^{-4}$ dependence.

The tree level relation (\ref{mass2}) yields then a lower bound for the 
lightest dual gauge boson mass:
\begin{equation}
M_1^{physical} \geq 3000\mbox{ TeV}.
\label{M1phys}
\end{equation}
Here we have used for the dual couplings $\tilde{g}_3$ and $\tilde{g}_1$ 
the Dirac conditions (\ref{couplings}) where we have taken the running
couplings of QED and QCD at the scale of the dual symmetry breaking 
($\zeta z$).  This last assumption is something which can be questioned 
and a second possibility, less conservative, is to take the QED and QCD 
couplings at a lower scale given by the energies relevant in the process, 
in which case the bound would be less severe.

\subsection{Air Showers beyond the GZK Cut-off}

As explained in the introduction, the suggestion that air showers with
energy higher than the GZK cut-off \cite{greitsemin} at $5\times 10^{19}$eV
are due to neutrinos having acquired a strong interaction from dual gauge
boson exchange \cite{airshower1,airshower2} puts a rough upper bound on 
the dual gauge boson mass.  At low energy, the effects due to dual gauge 
boson exchanges are suppressed generically by a factor $(\tilde{g}_{eff} 
\omega/M)^{-4} \sim (\omega /\zeta z)^{-4}$, where $\omega$ is the C.M. 
energy and $M$ the mass of the lightest dual gauge boson.  Therefore, for 
neutrinos to acquire a strong interaction from this dual gauge boson 
exchange, the C.M. energy should be such as to give $\omega/\zeta z$ of 
order unity.  Now a neutrino with incoming energy $10^{20}$eV impinging 
on a nucleon in an air nucleus at rest in the atmosphere has around 500 
TeV C.M. energy.  Since a neutrino is assumed at that sort of energies to 
produce air showers, its cross section with the air nucleus has to be 
hadronic and its interaction therefore strong.  From this, one can deduce 
then the following rough upper bound:
\begin{equation}
\zeta z \leq 500\ \rm{TeV}.
\label{ubound}
\end{equation}
It is clear that this bound is rather crude since it is arguable whether
$\tilde{g}_{eff} \omega/M$ need be as large as unity for the cross section
to be hadronic, nor is it obvious that it is the C.M. energy for a
neutrino-nucleon collision that should be taken rather than that for a
neutrino-nucleus collision.  It is nevertheless interesting that provided
one accepts that air showers beyond the GZK cut-off are neutrino induced
then one does obtain an upper bound for the scale parameter $\zeta z$
which is not far from the lower bound deduced above from the $K_L - K_S$
mass difference.  This is the only upper bound on the scale parameter 
that we are aware of at present.

In the literature, there has appeared a claim \cite{halzen} that the 
suggested neutrino explanation for post-GZK air showers does not work.
We do not think this claim is valid.  The arguments presented there,
leading to their conclusion that the neutrino cross section will be too small,
are based only on first order perturbation theory which is far from
adequate for dealing with the problem of (soft) hadronic cross sections 
addressed here which is a highly nonperturbative phenomenon.  Indeed,
with the same arguments, one would not obtain hadronic-sized cross 
sections even for proton-proton collisions.  It is said in that paper 
that its conclusion is a consequence of $s$-wave unitarity but no 
justification nor reference for this statement is given\footnote{One 
of us (CHM) thanks Francis Halzen for a series of email exchanges on 
this question, which has not, unfortunately, despite much effort on 
our part, succeeded in clarifying the basis of their assertion that 
$s$-wave unitarity implies the claim they made.}.  Our own searches in 
the literature also have not found any reference to that result.
We are sceptical that 
$s$-wave unitarity can give serious bounds on high energy hadron cross 
sections which involve many partial waves.  Even from full ($s$-channel) 
unitarity, apart from the geometric constraints already dealt with in
\cite{airshower1}, the bounds on the rate of increase of the cross
section  that we are aware of are asymptotic bounds which are not 
relevant for the case in hand where there are still new thresholds 
opening\footnote{We are deeply grateful to Andr\'e Martin for 
correspondence and for his expert advice on this matter.}.  Our 
conclusion is thus that the objection raised in \cite{halzen} is 
ill-founded and in no way affects the feasibility of the proposal 
\cite{airshower1,airshower2}.  Nevertheless, of course, the proposal 
must be subjected to much more stringent tests than it has so far 
undergone before it can be taken seriously.  Indeed, apart from the 
proposed tests in air shower physics itself as given in \cite{airshower1}, 
the best tests suggested are in the low energy FCNC effects considered 
in this paper and detailed below.

\subsection{Rare $K$ Decays}

Taking into account the quark content of the $K$-meson ($\bar{s}d$),
one finds that the piece of the effective Lagrangian (\ref{laeff1}) for 
semileptonic $K$ decays is:
\begin{equation}
L^{D,L}_{eff}=
\frac{1}{ \zeta^2 z^2} 
|f^{D,L}_{2,3;\alpha,\beta}|
\left( \bar{s}_L \gamma^\mu d_L \right)
\left( \bar{l}_{\alpha,L} \gamma_\mu l_{\beta,L} \right)
\label{kaons}
\end{equation}
where $l_{\alpha}(l_{\beta})$ stand for two different leptons of the same 
type.  To minimise the errors from uncertainties in the hadron structure 
we take quotients between the rare and Standard Model-allowed processes
which contain the same hadronic matrix elements.  The result then depend
on the ratio between the v.e.v.'s of the conventional (electroweak) and 
the new (dual colour) Higgs fields, as well as the group factors from the 
fermionic sector (\ref{groupfactor1}).  For instance, for $K^+$ rare 
decays one has:
\begin{equation}
\frac
{Br \left(K^+ \rightarrow \pi^+ l_{\alpha} l_{\beta} \right)}
{Br \left(K^+ \rightarrow \pi^0 \nu_{\mu} \mu^+ \right)}
=
|f^{D,L}_{2,3;\alpha,\beta}|^2 
\left( \frac{v}{\zeta \, \, z} \right)^4 \frac{2}{\sin^2 \theta_c},
\nonumber \\
\label{kplus}
\end{equation}
where $v=\frac{0.246}{\sqrt{2}}$ TeV and $\theta_c$ is the Cabibbo angle, 
$\sin \theta_c=0.23$.

Similarly, for the leptonic decays of the neutral $K$-mesons, one can 
write:
\begin{equation}
\frac
{\Gamma \left(K^0_{L(S)} \rightarrow l_\alpha l_\beta \right)}
{\Gamma \left(K^+ \rightarrow \mu^+ \nu_\mu \right)}
= |f^{D,L}_{2,3;\alpha,\beta}|^2
  \left( \frac{v}{\zeta \, \, z} \right)^4 \frac{1}{\sin^2 \theta_c},
\label{kls}
\end{equation}
from which, given the total widths of $K_L$ and $K_S$ and the width of
$K^+ \rightarrow \mu^+ \nu_\mu$ measured in experiment, one can easily 
calculate the branching ratios of the various leptonic modes of $K_L$ 
and $K_S$.

\begin{table}

\begin{eqnarray*}
\begin{array}{||l|l|l||}
\hline \hline
  & Theoretical \, \, Estimate & Experimental \, \, Limit \\
\hline
Br(K^+ \rightarrow \pi^+ e^+ e^-)  &  4 \times 10^{-15}   &
2.7 \times 10^{-7}  \\
Br(K^+ \rightarrow \pi^+ \mu^+ \mu^-)  & 2 \times 10^{-15}    &
2.3 \times 10^{-7}  \\
Br(K^+ \rightarrow \pi^+ e^+ \mu^-)  &  2 \times 10^{-15}   &
7 \times 10^{-9}  \\
Br(K^+ \rightarrow \pi^+ e^- \mu^+)  &  2 \times 10^{-15}  &
2.1 \times 10^{-10}   \\
Br(K^+ \rightarrow \pi^+ \nu {\bar \nu})  &  2 \times 10^{-14}   &
2.4 \times 10^{-9}  \\
Br(K_L \rightarrow e^+ e^-)  & 2 \times 10^{-13}   &
4.1 \times 10^{-11}  \\
Br(K_L \rightarrow \mu^+ \mu^-)  & 7 \times 10^{-14}    &
 7.2 \times 10^{-9} \\
 Br(K_L \rightarrow  e^{\pm} \mu^{\mp})  &  1 \times 10^{-13}    &
 3.3 \times 10^{-11} \\
 Br(K_S \rightarrow \mu^+ \mu^-)  &  1 \times 10^{-16}   &
 3.2 \times 10^{-7}\\
 Br(K_S \rightarrow  e^+ e^-)  &  3 \times 10^{-16}   &
2.8 \times 10^{-6} \\
 Br(K_S \rightarrow e^\pm \mu^\mp) & 2 \times 10^{-16} & ? \\
\hline \hline
\end{array}
\end{eqnarray*}
\caption{Branching ratios for semileptonic and leptonic rare $K$ decays. 
The two columns show respectively the tree-level predictions of the 
Dualized Standard Model and the present experimental limits.  The lowest 
v.e.v. $\zeta z$ of the Higgs fields is taken at 400 TeV.}
\end{table}

In Table 1 we have collected the tree-level predictions of the Dualized 
Standard Model for the rare semileptonic and leptonic decay modes of $K_L$, 
$K_S$ and $K^+$, and compared them with the limits reached by present 
experiments.  From the table we see that for the two modes $K_L \rightarrow 
e^\pm \mu^\mp$ and $K_L \rightarrow e^+ e^-$, the predicted branching ratios 
are roughly a couple of orders of magnitude down from the present experimental 
limits and may be accessible in the near future.  The other modes
appear out of 
reach for some time.

\subsection{Mass Difference in the Neutral $D$-Meson and $B$-Meson
Systems}

Using the lower bound on the Higgs v.e.v. $\zeta z$ obtained from the mass 
difference of neutral $K$'s, one can deduce lower bounds also on the mass
differences in the neutral $D$ and $B$ systems.

To study the $D$ system we proceed as before.  The relevant piece in 
the effective Lagrangian (\ref{laeff1}) is:
\begin{equation}
L^{U,U}_{eff}=
\frac{1}{2 \zeta^2 z^2} 
|f^{U,U}_{2,3;2,3}|
\left( \bar{c}_L \gamma^\mu u_L \right)
\left( \bar{c_{L}} \gamma_\mu u_{L} \right),
\label{massd0}
\end{equation}
and the contribution of the dual gauge boson to the mass splitting is:
\begin{equation}
{\Delta m_D} = \frac{m_D}{\zeta^2 z^2} \frac{f_D^2}{3} |f^{U,U}_{2,3;2,3}|.
\end{equation}
Taking the values $f_D^2=10^{-8}\mbox{ TeV}^2$ for the $D$-meson coupling and
$m_D=1.865$ GeV for the  mass we have:
\begin{equation}
 {\Delta m_D} = 5 \times 10^{-12}\mbox{ MeV}.
\end{equation}
This is one and a half orders of magnitude below the present experimental 
limits ${\Delta m_D}\leq 1.4 \times 10^{-10}$ MeV and could be accessible to
planned experiments in the near future.
 
Applying the same procedure to the mass-splitting between the neutral 
$B$-mesons, one finds that the contribution from dual gauge boson exchange
is 6 orders of magnitude smaller than that from the Standard Model as 
can be seen by comparing the coupling in $L_{eff}$ (\ref{laeff1}) to the 
current $\bar{b}_L \gamma^\mu s_L$:
\begin{equation}
\frac{|f^{D,D}_{2,1;2,1}|}{\zeta^2 z^2} = 5 \times 10^{-10}
\, \, \, {\rm TeV}^{-2},
\end{equation}
and its counterpart in the Standard Model:
\begin{equation}
\sin^4 \theta_c \left( \frac{m_t}{37\ {\rm GeV}} \right)^2
\frac{\alpha_{EM}}{4 \pi v^2} 
\approx 10^{-3} \, \, \, {\rm TeV}^{-2}.
\end{equation}
The small contribution coming from the dual sector, apart from the large 
value of the intermediate dual boson mass, is due to the smallness of 
the dual colour group factor which here involves the small element (1,3) (the
torsion term, according to \cite{features}) of the mixing matrix 
(\ref{mixingf}).

\subsection{Rare $D$ and $B$ Decays}

We perform here the same exercise for the semileptonic decays of mesons 
which contain the quarks $c$ and $b$ as we did for $K$.

$D$-meson rare decays, such as $D \rightarrow \pi l_{\alpha} l_{\beta} $,
are controlled by the term of the effective Lagrangian:
\begin{equation}
L^{U,L}_{eff}=
\frac{1}{2 \zeta^2 z^2} 
|f^{U,L}_{2,3;\alpha,\beta}|
\left( \bar{c}_L \gamma^\mu u_L \right)
\left( \bar{l}_{\alpha,L} \gamma_\mu l_{\beta,L} \right).
\label{leffd}
\end{equation}
Therefore, in the spectator quark model, which roughly predicts the correct
magnitudes for $D$-meson lifetimes, the decay width for the process is 
given by:
\begin{equation} 
\Gamma \left( D^+ \rightarrow \pi^+ l_{\alpha} l_{\beta} \right)
=
|f^{U,L}_{2,3;\alpha,\beta}|^2 
\frac{m_c^5}{192 \pi^3} \frac{1}{8 \zeta^4 z^4},
\nonumber
\end{equation}
which, for a total width of $\Gamma=9.5 \times 10^{11} s^{-1}$ and a mass 
for the $c$ quark $m_c=1.3$ GeV gives an upper bound for the branching 
ratio for semileptonic rare decays:
\begin{equation}
Br \left(D^+ \rightarrow \pi^+ l_{\alpha} l_{\beta} \right) =
|f^{U,L}_{2,3;\alpha,\beta}|^2  \,\,\, 5 \times 10^{-15}.
\end{equation}

The analogous expression for leptonic decays of the neutral $D$-meson is:
\begin{equation}
Br \left(D^0 \rightarrow l_\alpha l_\beta \right) =
|f^{U,L}_{2,3;\alpha,\beta}|^2  \,\,\, 2 \times 10^{-15},
\end{equation}
where the only change comes from the difference in total widths between
the charged and neutral mesons.

In Table 2 we list the numerical values for the various modes.  They are
far from the present experimental bounds and seem inaccessible in the
near future.

\begin{table}

\begin{eqnarray*}
\begin{array}{||l|l|l||}
\hline \hline
  & Theoretical \, \, Estimate & Experimental \, \, Limit \\
\hline
Br(D^+ \rightarrow \pi^+ e^+ e^-)  &  3 \times 10^{-16}  &
6.6 \times 10^{-5}   \\
Br(D^+ \rightarrow \pi^+ \mu^+ \mu^-)  &  1 \times 10^{-16}   &
1.8 \times 10^{-5}  \\
Br(D^+ \rightarrow \pi^+ e^\pm \mu^\mp)  & 2 \times 10^{-16}    &
3.3 \times 10^{-3}  \\
Br(D^0 \rightarrow e^+ e^-) & 1 \times 10^{-16} & 1.3 \times 10^{-5} \\
Br(D^0 \rightarrow \mu^+ \mu^-) & 4 \times 10^{-17} & 7.6 \times 10^{-6} \\
Br(D^0 \rightarrow e^\pm \mu^\mp) & 6 \times 10^{-17} & 1.9 \times 10^{-5} \\
\hline \hline
\end{array}
\end{eqnarray*}
\caption{Branching ratios for semileptonic and leptonic rare $D$ decays. 
The two columns show respectively the tree-level predictions of the Dualized 
Standard Model and the present experimental limits.  The v.e.v.\ of the 
Higgs field taken at 400 TeV.}
\end{table}

The same calculation for B-meson decays gives the branching ratio for the 
modes with the $d$ quark in the final state as:
\begin{equation}
Br \left(B^+ \rightarrow \pi^+ l_{\alpha} l_{\beta} \right) =
|f^{U,L}_{1,3;\alpha,\beta}|^2  \,\,\, 2 \times 10^{-12}
\end{equation}
and with the $s$ quark in the final state as:
\begin{equation}
Br \left(B^+ \rightarrow K^+ l_{\alpha} l_{\beta} \right) =
|f^{U,L}_{1,2;\alpha,\beta}|^2  \,\,\, 2 \times 10^{-12}.
\end{equation}

As in the case of the neutral $B$ mass difference, the (1,3) element
of the mixing mass matrix reduces the theoretical values by 6 orders
of magnitude.  Thus, the present experimental limits in this case (being 
in the interval $10^{-3}-10^{-5}$) are far from the values predicted in the
Dualized Standard Model.

\section{Remarks}

We have evaluated above the predictions of the Dualized Standard Model
in tree-level approximation to a number of flavour-changing effects based
on bounds to the dual gauge boson mass scale obtained from $K_L - K_S$
mass difference and from extremely high energy air showers.  These, though
interesting and hopefully useful to experimenters planning experiments,
are subject to a number of reservations.

First, as has already been made clear, the estimate for the mass scale
$\zeta z \sim 400$ TeV is rather crude, especially the upper bound from
post-GZK air showers, even if the proposed neutrino explanation for these 
is accepted as valid.  Given that the FCNC effects we seek are unfortunately
dependent on this scale to the fourth power, the absolute normalization 
of our predictions can easily be wrong by say an order of magnitude.
Ultimately, the best way to determine the scale would be to measure
the actual magnitude of one of the FCNC effects studied in this paper.

The relative sizes of the effects, on the other hand, depend at tree-level
only on the rotation matrices $S$ of (\ref{mixingf}) which are fitted
quite sensitively to the CKM matrices for quarks and seem in agreement
with experiment for leptons.  They are thus considered more reliable.
Given that the fits to the lowest generation quark and lepton masses 
\cite{ckm} are not perfect, future adjustments are probably necessary
but are felt unlikely to affect the matrices $S$ too drastically.

Next, all the above analyses have been performed only at tree-level.  
However, the dual gauge couplings being large by virtue of the Dirac 
quantization condition (\ref{couplings}) one has to bear in mind the 
possible effects of higher order terms.  Assuming still the tree-level 
spectrum for the dual gauge boson states, and using the effective Lagrangian 
formalism above, we have calculated some higher order corrections, in 
particular the most important ones coming from the box diagram with the 
exchange of two gauge dual bosons. This contribution changes the effective 
coupling by a factor:
\begin{equation}
\frac{\tilde{g}_3^2}{4 \pi^2} \frac{1}{k_1} \frac{3}{16} n_L^A n_L^B ,
\label{2ndorder}
\end{equation}
which directly depends on the hypercharge of the fermion-types involved,
following \cite{chantsou2}, we take $n_L^A=-1$ for all of them. 
Because of the unitarity properties of the mixing matrices (\ref{mixingf})
there are no additional group factors of the kind (\ref{groupfactor1}).  
The factor in (\ref{2ndorder}) can be estimated relating the coupling
constant $\frac{\tilde{g}_3^2}{4 \pi}$ to the QCD running coupling
constant (\ref{couplings}). At the energies relevant to the meson decays
considered here, it is of the order of unity and, as 
a result, the correction due to the box diagram is of order 20\% of the 
tree-level contribution.  If so, it is still not too bad.

On the other hand, the effect of higher order corrections to the tree-level 
dual gauge boson spectrum itself is more worrisome and harder to handle.  
At tree-level, the dual gauge boson masses are given in terms of just 
the v.e.v.'s of the Higgs fields, and from the values of these deduced from
our earlier fits to the CKM matrix \cite{ckm} we identified the boson 
with the lowest mass which in the end dominates over all others in giving 
FCNC effects.  This conclusion arose from the diagonalization of the 
matrix (\ref{mass1}) with its rather special property, namely that for 
$y = z = 0$ the matrix factorizes, hence giving it two near-zero eigenvalues 
for $x \gg y \gg z$.  We do not know whether this property will be preserved
by loop corrections, and if not then our predictions for FCNC effects may
be greatly altered.  Let us consider as an illustration a scenario at 
the opposite extreme to that we have considered, namely when all dual 
gauge bosons have roughly equal mass.  In that case, the sum in (\ref{lag3}) 
runs over the whole range with all dual gauge bosons contributing at the 
same level to flavour-changing processes.  One gets then, for processes 
with change of flavour in at least one of the fermion vertices:
\begin{equation}
f^{A,B}_{\alpha,\beta;\alpha',\beta'} =
\left( S^{A \,\, \dagger} S^{B} \right)_{\alpha,\beta'}
\left( S^{B \,\, \dagger} S^{A} \right)_{\alpha',\beta} .
\label{groupfactor2}
\end{equation}
Notice that, between fermions of the same type, i.e.\ $A=B$, this factor 
vanishes.  There will then be no contribution to $K_L - K_S$ mass 
difference and the most stringent lower bound on dual gauge boson mass
will arise instead from the $K$ decay $K_L \rightarrow e^+ e^-$.  One
obtains then from the present experimental limit on this decay a bound 
on the dual gauge bosons mass which translates to a Higgs v.e.v. of only 
$20$ TeV as compared with the $400$ TeV obtained before, giving upper 
bounds for rare decay processes some $10^4$ times higher than the values 
given in Table 1. 

However, it should be stressed that such an extreme situation need not 
happen, nor do we have to expect necessarily large loop corrections in
spite of the large dual gauge couplings $\tilde{g}_i$.  As emphazised
in \cite{chantsou2}, the dual gauge bosons do not represent a different
degree of freedom from the ordinary gauge bosons so that corrections
from dual gauge boson loops really make no sense as an addition to ordinary
gauge bosons loops and these latter need not be large.  At present,
however, we have to admit we do not know how to proceed beyond the
tree-level and await a better understanding of the phenomenon called
`metamorphosis' in \cite{chantsou2} which is under investigation.

Despite the limitations outlined above, the analysis presented is interesting
in giving for, we believe, the first time, a full calculation of all FCNC
effects in terms of only one scale parameter, and even this is constrained 
by experiment if the neutrino explanation of post-GZK air showers 
\cite{airshower1,airshower2} is to be believed.  This is made possible
by the ability of the Dualized Standard Model to identify the rotation
matrices $S$ in (\ref{mixingf}) relating the gauge and physical fermion 
states of which we have some confidence.  Even if the tree-level spectrum 
for the dual gauge bosons turns out to be invalid because of loop corrections, 
one can imagine using in future some of the FCNC effects to fix the 
spectrum and then apply the $S$ matrices to deduce the rest of FCNC
effects.  At present, however, the tree-level is the best that we can
do, but we still think it should be useful as a guide.

The various FCNC effects investigated above mostly give estimates a
long way below the present experimental bounds.  Of the few that appear
possibly accessible to experiment in the near future, the 2 decays
$K_L \rightarrow e^+e^-$ and $K_L \rightarrow \mu^+\mu^-$ can occur 
also by second order weak interaction, with  branching ratios estimated
by a recent paper \cite{who} to be respectively around $9 \times 10^{-12}$
and $1.2 \times 10^{-9}$, which are above the estimates obtained here
by dual gauge boson exchange, and are not therefore useful for testing
the latter.  The mode $K_L \rightarrow e^\pm \mu^\mp$, on the other 
hand, which also gives an estimate close to the present experimental 
limit, cannot proceed by second order weak interaction if neutrinos
do not mix.  Even if neutrinos do mix, the contribution from second 
order weak interaction will be suppressed by the square of the mixing 
angle which will make it comparable to the estimate obtained here from
dual gauge boson exchange.  We think, therefore, that this decay is a 
particularly good angle to attack, since a postive result here would
give a confirmation either of neutrino mixing or of the dual gauge
boson exchange process we have here advocated.  Another effect worth
noting, apart of course from the $K_L - K_S$ mass difference, is the 
mass difference of the neutral $D$-mesons, for which our estimate is 
only an order of magnitude below the present experimental bound.

\vspace{.5cm}
\noindent{\large {\bf Acknowledgment}}\\

One of us (JB) wishes to thank A. Santamaria and F. Botella for very 
interesting discussions and to acknowledge support from the Spanish 
Government on contract no. CICYT AEN 97-1718, while another (JP) is 
grateful to the Studientstiftung d.d. Volkes and the Burton Senior 
Scholarship of Oriel College, Oxford, for financial support.

\end{document}